# Freedericksz-like positional transition of a micro-droplet suspended in a nematic cell

Ke Xiao, Xi Chen, and Chen-Xu Wu*
*Department of Physics, College of Physical Science and Technology, Xiamen University, Xiamen 361005, Peoples Republic of China*

In this paper, a Freedericksz-like positional transition is found for a spherical micro-droplet suspended in a nematic liquid crystal cell in the presence of an external electric field. Based on the numerical calculation of elastic energy using Green function method, the equilibrium position of micro-droplet is decided through a competition between the buoyant force and the effective force built by the elastic energy gradient existing inside the nematic liquid crystal(NLC) cell. It is shown that the elastic energy dominates the kinetics of micro-droplet until the external field applied reaches a critical value large enough to flatten the elastic energy contour in the central region, which enables the asymmetric buoyant force to drive the liquid droplet abruptly from the cell midplane to a new equilibrium position. It is also found that such a threshold value of external field, which triggers positional transition, depends on thickness $L$ and Frank elastic constant $K$, in a Freedericksz-like manner, but multiplied by a factor of $3\sqrt{\pi}$. An explicit formula proposed for the critical electric field agrees extremely well with the numerical calculation.

Behaviors of immiscible liquid, solid or gas microphase suspensions in a nematic liquid crystal(NLC) cell are of considerable interest due to their promising practical applications in new display devices and materials[1–3], triggered release and reporting of microcargo[4], and biological detectors[5, 6]. So far lots of efforts have been made to investigate behaviors of microparticles suspended in confined NLCs by means of experiment and numerical modelling[7–16]. Physical phenomena such as levitation, lift, bidirectional motion and aggregation of colloids induced by the effect of electric field were also studied through experiments and computer simulations[17, 18]. Diverse methods and techniques have been developed to measure the interaction force between particles in NLC in a direct manner[7, 19–22].

The interaction force of spherical particles suspended in NLC is associated not only with interparticle distance and geological confinement[8], but also with shape of particles which plays a crucial role in pair interaction and aggregation behaviors[12]. In most cases the inclusion of particles into a NLC cell tends to create LC alignment singularities around the suspended substances, which in general are determined by surface anchoring conditions, particle size, boundary conditions, and external fields[12, 23–26]. It has been widely confirmed and accepted that when a spherical particle is immersed in NLC, there are three possible types of defect configurations[27–29]. Dipole and quadrupolar configurations are usually seen around a spherical particle with strong vertical surface anchoring, whereas boojum defect is formed by a micro-sphere with tangential surface anchoring. Through experimental observations it has been found that, when an external field is applied, there exists a transition between elastic dipole and quadrupolar configuration, which depends on particle size and surface anchoring strength[30–32].

On the theoretical side, recently S. B. Chernyshuk and coauthors studied the interaction between colloidal particles in NLCs near one wall and in a nematic cell with or without external field by using method of Green function, and obtained general formulae for interaction energy between colloidal particles[33–35]. Although interactions of two particles in a NLC are very well understood and the particle-wall interaction has been widely observed experimentally for a single particle immersed in a nematic cell[34, 36, 37], the properties of a single particle in a uniform NLC cell in the presence of an external electric field theoretically have not been fully addressed.

The system we consider consists of a micro-droplet of radius $r$ suspended in a NLC cell in the presence of an external electric field, as illustrated in Fig. 1. Experiments showed that typically such a liquid droplet tends to generate a companion hyperbolic hedgehog rather than a disclination ring[29], and thus it is appropriate to consider the micro-droplet and its surrounding LC alignments as a dipole configuration. With one Frank constant approximation, the effective elastic energy for the system is given by[35]

$$U_e = K \int d^3x \left[ \frac{(\nabla n_\mu)^2}{2} - \frac{k^2}{2}(\mathbf{en})^2 - 4\pi P(\mathbf{x})\partial_\mu n_\mu - 4\pi C(\mathbf{x})\partial_z \partial_\mu n_\mu \right], \tag{1}$$

where $K$ is the Frank constant, $n_\mu$ ($\mu = x, y$) represents the components of director field $\mathbf{n}$ perpendicular to $\mathbf{n}_0$, and $k^2 = (4\pi K)^{-1}\Delta\varepsilon E^2$ with $\Delta\varepsilon = \varepsilon_\parallel - \varepsilon_\perp$ the dielectric anisotropy of the NLC. $P(\mathbf{x})$ and $C(\mathbf{x})$ denote the dipole- and the quadrupole-moment densities respectively. For simplicity we choose the coordinate $z$ axis along the normal direction of the two cell walls on which LC molecules are homeotropically anchored. Given these, when an electric field along $z$ axis is applied, we have the Euler-Lagrange equations[35]

$$\Delta n_\mu - k^2 n_\mu = 4\pi[\partial_\mu P(\mathbf{x}) - \partial_z \partial_\mu C(\mathbf{x})]. \tag{2}$$

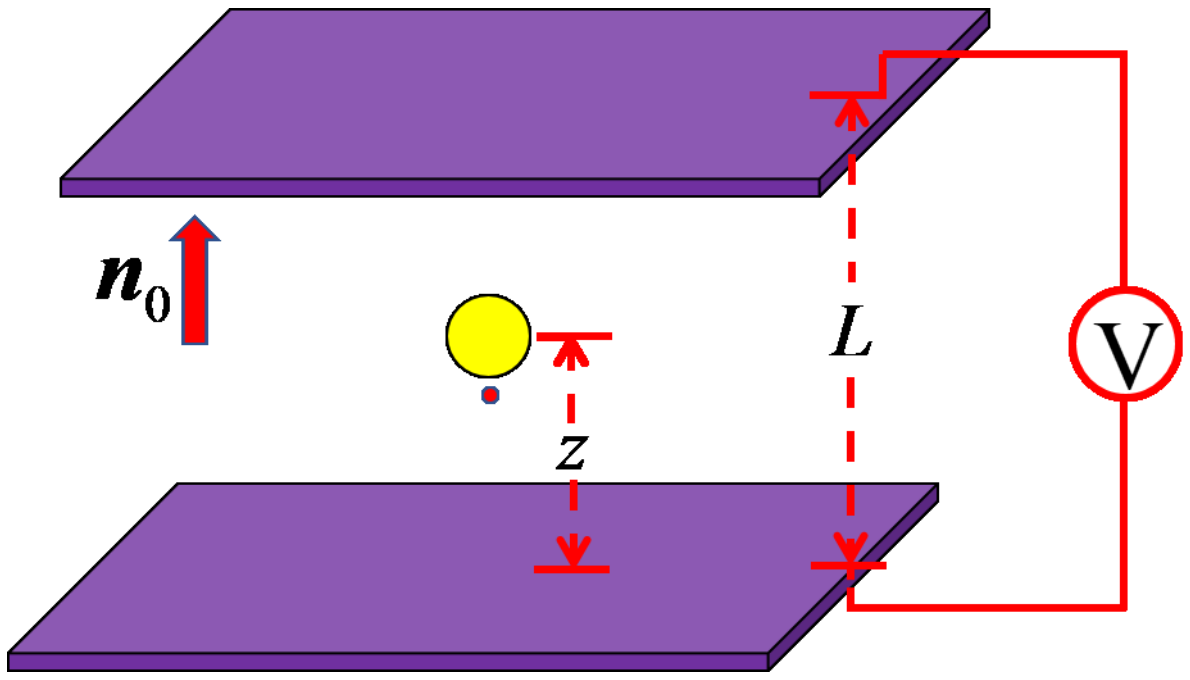

FIG. 1. (Color online) Sketch of a micro-droplet of radius $r$ suspended in a nematic cell with $L$-thick spacer in the presence of an external field.

With Dirichlet boundary conditions $n_\mu|_{(z=0)} = n_\mu|_{(z=L)} = 0$ for homeotropic anchoring on the two walls, the solution

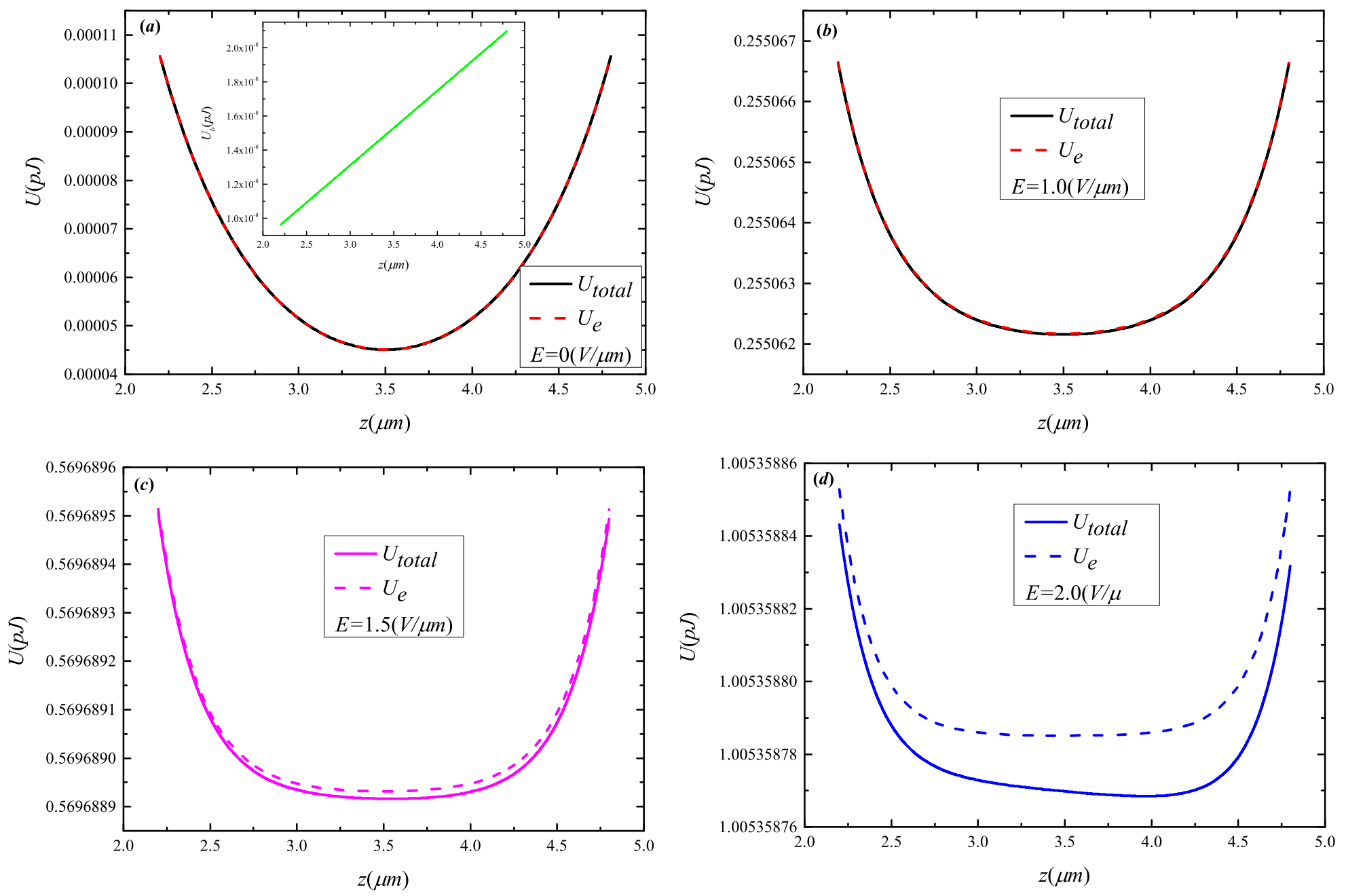

FIG. 2. (Color online) Elastic energy and total energy as a function of micro-droplet position for different electric field(0, 1.0, 1.5 and 2.0$V/\mu m$). Here we set the radius of micro-droplet and cell thickness as 2.2$\mu m$ and 7$\mu m$, respectively.

of Euler-Lagrange equation can be written as[35]

$$n_\mu(\mathbf{x}) = \int_V d^3\mathbf{x}' G_\mu(\mathbf{x},\mathbf{x}')[-\partial'_\mu P(\mathbf{x}') + \partial'_\mu \partial'_z C(\mathbf{x}')], \tag{3}$$

where $G_\mu$ is the Green function given by[35]

$$G_\mu(\mathbf{x},\mathbf{x}') = \frac{4}{L}\sum_{n=1}^{\infty}\sum_{m=-\infty}^{\infty} e^{im(\varphi-\varphi')}\sin\frac{n\pi z}{L}\sin\frac{n\pi z'}{L} I_m(\lambda_n\rho_<)K_m(\lambda n\rho_>). \tag{4}$$

Here $I_m$ and $K_m$ are modified Bessel functions, and $\lambda_n = [(\frac{n\pi}{L})^2 + \frac{\Delta\varepsilon E^2}{4\pi K}]^{\frac{1}{2}}$. Then substituting $H_\mu(\mathbf{x},\mathbf{x}') = G_\mu(\mathbf{x},\mathbf{x}') - \frac{1}{\mathbf{x}-\mathbf{x}'}$ into the self-energy defined in terms of Green function in Ref.[35], we obtain the elastic energy $U_e$ for an NLC cell with a micro-droplet suspended in the presence of an electric field. Besides the elastic energy, the gravitational potential $U_g$ due to buoyant force should be considered as well, leading to a total energy written as

$$U_{total} = U_e + U_g = -2\pi K p^2 \left[ -\frac{4}{L} \sum_{n=1}^{\infty} \lambda_n^2 \sin^2(\frac{n\pi z}{L}) K_0(\lambda_n \rho) + \frac{1}{\rho^3} \right]_{\rho \to 0} - \frac{4}{3}\pi r^3 (\rho_{LC} - \rho_{aq}) g z, \quad (5)$$

where $r$ is the radius of micro-droplet, $p = 2.04r^2$ is the magnitudes of the dipole moment, $\rho_{LC} - \rho_{aq}$ is the density difference between liquid crystal and micro-droplet, $g = 9.8m/s^2$ is the gravitational acceleration, and $z$ denotes the position of micro-droplet.

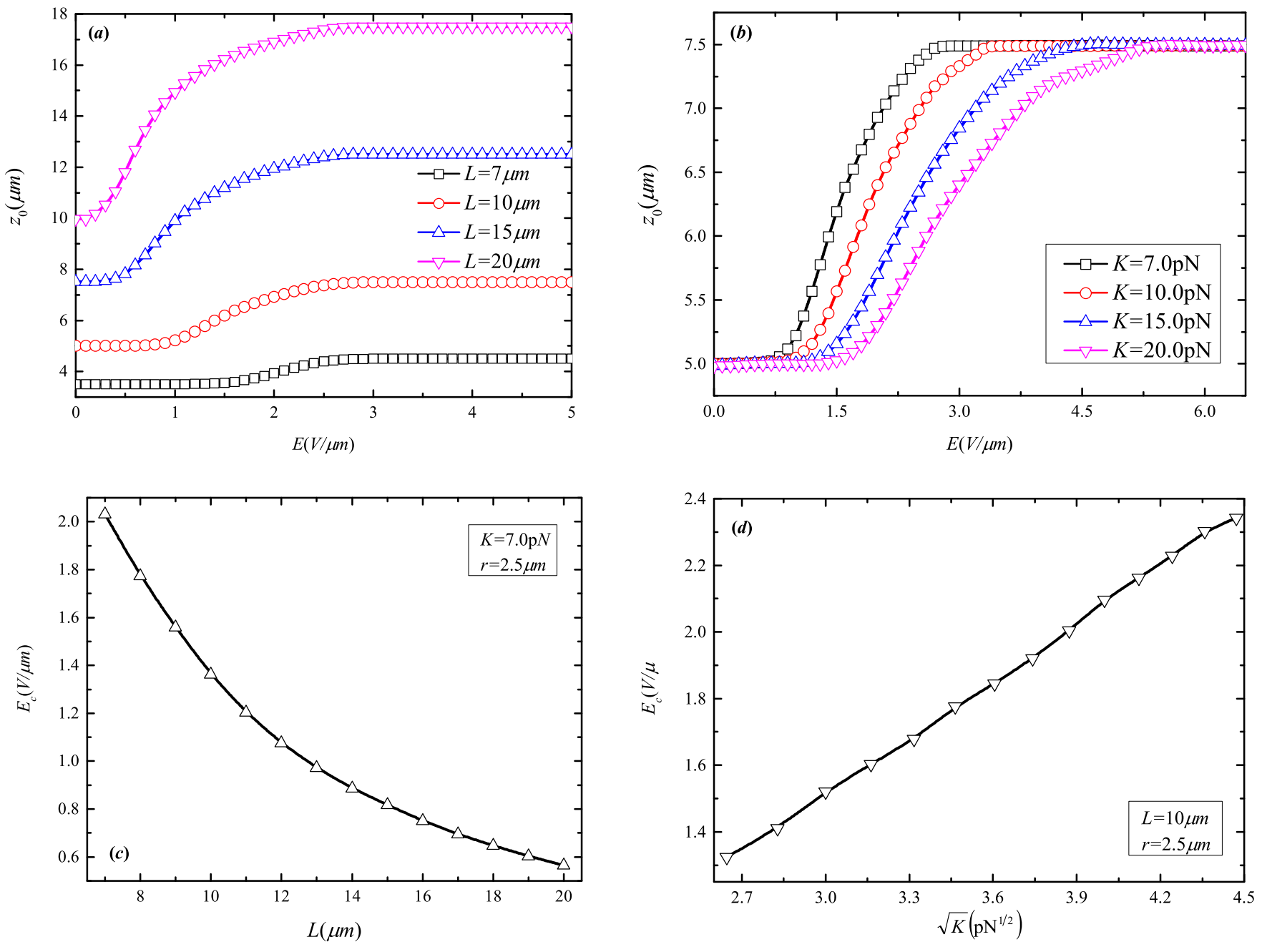

FIG. 3. (Color online) Equilibrium position $z_0$ in response to electric field for different (a) cell thickness $K = 7.0\ pN$ and $r = 2.5\ \mu m$, and (b) Frank elastic constant ($L = 10\ \mu m$ and $r = 2.5\ \mu m$), showing a positional transition occurring at electric field threshold $E_c$, which depends on (c) cell thickness $L$ and (d) Frank elastic constant $\sqrt{K}$.

In the presence of small external field, the total energy given by Eq. (5) overlaps the elastic energy $U_e$ and remains symmetric, indicating that the interaction among liquid crystal (LC) molecules still dominates the system if the external field applied is not large enough to realign the LC molecules, especially in the region close to the midplane. Thus the contribution made by asymmetric gravitational potential is trivial and the micro-droplet in this case is still trapped within its mid-plane, as shown in Fig. 2(b) and Fig. 2(c). But as we increase the field applied, it tends to widen and flatten the bottom of the elastic potential well, and that by contrast enlarges the contribution made by the asymmetric buoyant force to the total energy. As a result, the buoyant force will drive the micro-droplet with ease from midplane to a new equilibrium position (Fig. 2(d)). It seems that interaction potential well around the midplane tends to be "ironed out" by the applied external field, creating a smooth fast lane for the liquid bubble to shift from midplane with ease, if driven by an asymmetric buoyant force.

Furthermore, in order to study the effect of cell thickness and Frank constant on the critical electric value, we plot the equilibrium position against the applied electric field for different cell thickness(7, 10, 15 and 20$\mu m$) and Frank elastic constant(7, 10, 15 and 20pN), as shown in Figs.3(a) and (b). It is found that a positional transition occurs when the external field applied reaches a threshold value. The thinner the cell thickness $L$ is and the larger the Frank elastic constant $K$ is, the larger the critical electric field is needed to trigger the positional transition, as shown in Fig.3(c) and (d). A more deeper investigation shows that the critical value of electric field is inversely proportional to $L$ and linearly proportional to $\sqrt{K}$, a Freedericksz-like behavior.

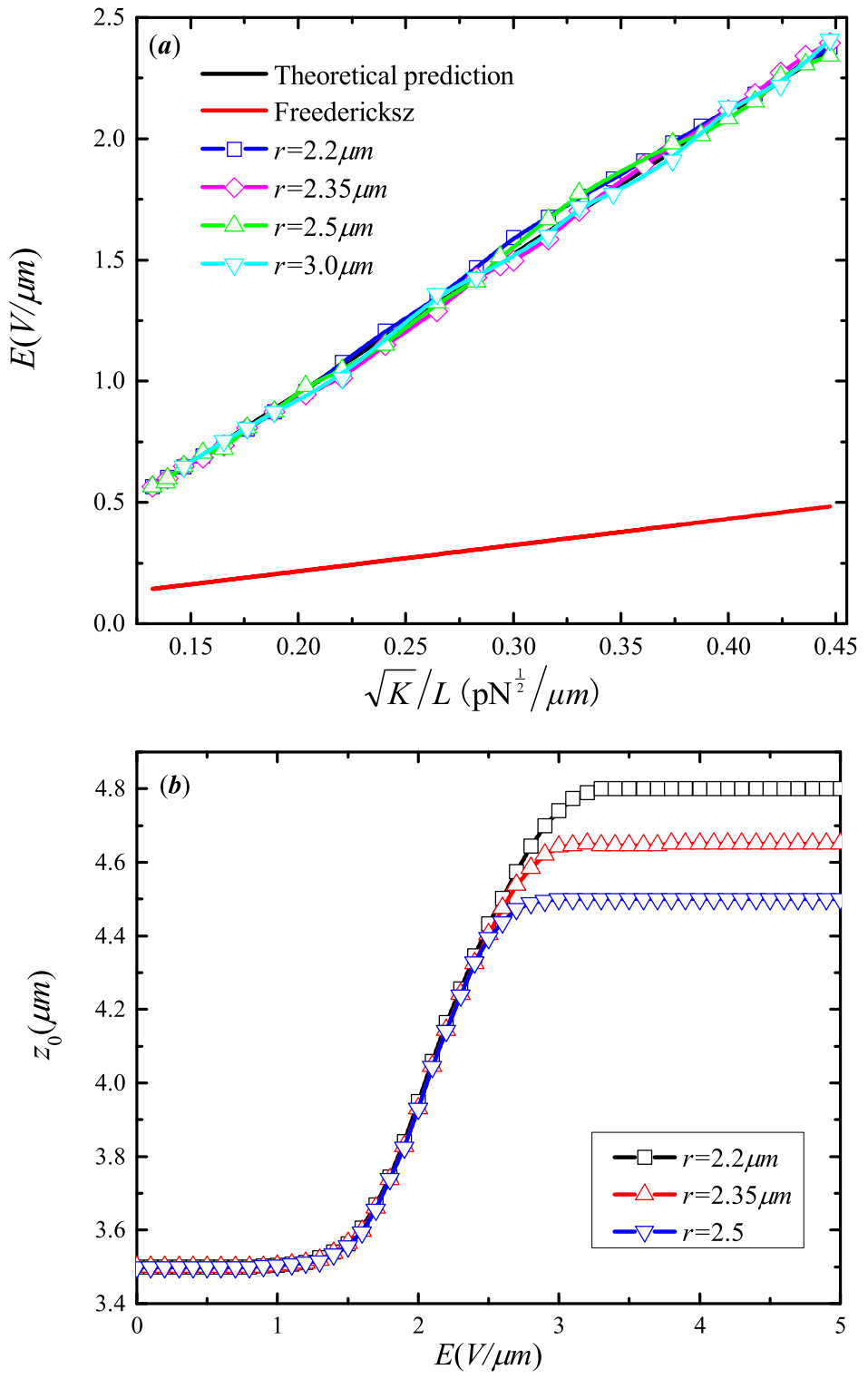

FIG. 4. (Color online) (a) The dependence of $E_c$ and $\sqrt{K}/L$ for different radii of micro-droplet ($2.2\mu m$, $2.35\mu m$, $2.5\mu m$ and $3.0\mu m$), obeying strictly a master curve given by theoretical prediction Eq. (6). (b) Equilibrium position $z_0$ for different radii of micro-droplet with $K = 7\ pN$ and $L = 7\ \mu m$, showing the same critical value $E_c$ of electric field triggering positional transition.

To gain more insight into the dynamic behaviors of the micro-droplet, we plot the threshold value against $\sqrt{K}/L$, so as to obtain a master curve, as shown in Fig. 4(a), where a Freedericksz curve (red) is also plotted. It is interesting to find that the critical electric field to trigger a positional transition for micro-droplet suspended in a NLC cell follows a Freedericksz-like linear master curve, yet with a different slope. In addition, one finds no or slight dependence of critical value on size of liquid droplet, as shown in Figs. 4(b).

Moreover, by comparing the numerical calculation results with the Freedericksz effect curve($\pi\sqrt{\frac{4\pi}{|\Delta\varepsilon|}}\frac{\sqrt{K}}{L}$) in Fig. 4(a), it is surprising to find that the slope difference between them is by a factor of $3\sqrt{\pi}$. More specifically, an explicit expression

$$E_c = 3\sqrt{\pi} * \text{Freedericksz effect} - \frac{1}{5} = 6\pi^2\sqrt{\frac{K}{|\Delta\varepsilon|L^2}} - \frac{1}{5} \tag{6}$$

for critical electric field can be proposed as a theoretical prediction. Such a prediction, as shown by straight line in Figs. 4(a), agrees very well for different radii of micro-droplet(2.2, 2.35, 2.5, and $3.0\mu m$). This once again verifies the conclusion that the critical electric field is independent of micro-droplet size. The reason might lie in that in the present theoretical model, the micro-droplet is treated as a dipole in the far field expansion approximation.

In summary, a positional transition is found for a micro-droplet suspended in a NLC cell in the presence of external electric field. The critical value of electric field that triggers such a transition is independent of droplet size in far-field approximation and obeys a Freedericksz-like dependence on cell thickness and Frank elastic constant in a master law by a factor of $3\sqrt{\pi}$. Such a theoretical prediction agrees very well with the numerical calculation results.

* cxwu@xmu.edu.cn

[1] B. Comiskey, J. Albert, H. Yoshizawa, and J. Jacobson, Nature **394**, 253 (1998).
[2] Z. Wang and J. Zhe, Chip **11**, 1280 (2011).
[3] I. I. Smalyukh, Annu. Rev. Condens. Matter Phys. **9**, 207 (2018).
[4] Y.-K. Kim, X. Wang, P. Mondkar, E. Bukusoglu, and N. Abbott, Nature **557**, 539 (2018).
[5] E. A. Nance, G. F. Woodworth, K. A. Sailor, T.-Y. Shih, Q. Xu, G. Swaminathan, D. Xiang, C. Eberhart, J. Hanes, Sci. Transl. Med. **4**, 149ra119 (2012).
[6] S. J. Woltman, G. D. Jay, and G. P. Crawford, Nat. Mater. **6**, 929 (2007).
[7] P. Poulin, V. Cabuil, and D. A. Weitz, Phys. Rev. Lett. **79**, 4862 (1997).
[8] M. Vilfan, N. Osterman, M. Čopič, M. Ravnik, S. Žumer, J. Kotar, D. Babič, and I. Poberaj, Phys. Rev. Lett. **101**, 237801 (2008).
[9] U. Ognysta, A. Nych, V. Nazarenko, I. Muševič, M. Škarabot, M. Ravnik, S. Žumer, I. Poberaj, and D. Babič, Phys. Rev. Lett. **100**, 217803 (2008).
[10] M. Škarabot, M. Ravnik, S. Žumer, U. Tkalec, I. Poberaj, D. Babič, N. Osterman, and I. Muševič, Phys. Rev. E **77**, 031705 (2008).
[11] A. V. Ryzhkova, M. Škarabot, and I. Muševič, Phys. Rev. E **91**, 042505 (2015).
[12] C. P. Lapointe, T. G. Mason, and I. I. Smalyukh, Science **326**, 1083 (2009).
[13] U. M. Ognysta, A. B. Nych, V. A. Uzunova, V. M. Pergamenschik, V. G. Nazarenko, M. Škarabot, and I. Muševič, Phys. Rev. E **83**, 041709 (2011).
[14] S.-J. Kim, B.-K. Lee, and J.-H. Kim, Liq. Cryst. **43**, 1589 (2016)
[15] D. Andrienko, M. Tasinkevych, P. Patrício, M. P. Allen, and M. M. Teloda Gama, Phys. Rev. E **68**, 051702 (2003).
[16] K. Izaki and Y. Kimura, Phys. Rev. E **87**, 062507 (2013).
[17] O. P. Pishnyak, S. Tang, J. R. Kelly, S. V. Shiyanovskii, and O. D. Lavrentovich, Phys. Rev. Lett. **99**, 127802 (2007).
[18] O. P. Pishnyak, S. V. Shiyanovskii, and O. D. Lavrentovich, J. Mol. Liq. **164**, 132 (2011).
[19] F. L. Calderon, T. Stora, O. Mondain Monval, P. Poulin, and J. Bibette, Phys. Rev. Lett. **72**, 2959 (1994).
[20] M. Yada, J. Yamamoto, and H. Yokoyama, Phys. Rev. Lett. **92**, 185501 (2004).
[21] K. Takahashi, M. Ichikawa, and Y. Kimura, Phys. Rev. E **77**, 020703(R) (2008).
[22] M. Škarabot, A. V. Ryzhkova, and I. Muševič, J. Mol. Liq. **267**, 384 (2018).
[23] H. Stark, Phys. Rep. **351**, 387 (2001).
[24] H. Stark, Phys. Rev. E **66**, 032701 (2002).
[25] Y. Wang, P. Zhang, and J. Z. Y. Chen, Phys. Rev. E **96**, 042702 (2017).
[26] X. Yao, H. Zhang, and J. Z. Y. Chen, Phys. Rev. E **97**, 052707 (2018).
[27] P. Poulin, H. Stark, T. C. Lubensky, and D. A. Weitz, Science **275**, 1770 (1997).
[28] P. Poulin and D. A. Weitz, Phys. Rev. E **57**, 626 (1998).
[29] T. C. Lubensky, D. Pettey, N. Currier, and H. Stark, Phys. Rev. E **57**, 610 (1998).
[30] R. W. Ruhwandl and E. M. Terentjev, Phys. Rev. E **54**, 5204 (1996).
[31] R. W. Ruhwandl and E. M. Terentjev, Phys. Rev. E **56**, 5561 (1997).
[32] J. C. Loudet and P. Poulin, Phys. Rev. Lett. **87**, 165503 (2001).
[33] S. B. Chernyshuk and B. I. Lev, Phys. Rev. E **81**, 041701 (2010).
[34] S. B. Chernyshuk and B. I. Lev, Phys. Rev. E **84**, 011707(2011).
[35] S. B. Chernyshuk, O. M. Tovkach, and B. I. Lev, Phys. Rev. E **85**, 011706 (2012).
[36] S.-J. Kim and J.-H. Kim, Soft Matter **10**, 2664 (2014).
[37] B.-K. Lee, S.-J. Kim, B. Lev, and J.-H. Kim, Phys. Rev. E **95**, 012709 (2017).